\documentclass{article}
\usepackage{algorithm}
\usepackage[noend]{algorithmic}
\usepackage{amssymb}
\usepackage{pst-tree,multirow,fullpage}
\usepackage{url}
\newtheorem{dfn}{Definition}

\newtheorem{theo}{Theorem}
\newenvironment{proof}{\paragraph{Proof.}}{$\Box$}

\begin{document}

\title{A Scalable Algorithm for\\ Minimal Unsatisfiable Core 
Extraction\thanks{This research was supported
by the Israel Science Foundation (grant no.\ 250/05).
The work of Alexander Nadel was carried out in partial fulfillment of the 
requirements for a Ph.D.}}

\author{Nachum Dershowitz\\
School of Computer Science\\ Tel Aviv University, Ramat Aviv, 
Israel\\
\url{nachumd@post.tau.ac.il}\\
\and Ziyad Hanna \\
Design Technology Solutions Group\\ Intel Corporation, Haifa, Israel\\
\url{ziyad.hanna@intel.com}\\
\and Alexander 
Nadel\\
School of Computer Science\\ Tel Aviv University, Ramat Aviv, 
Israel\\
\url{ale1@post.tau.ac.il}\\
Design Technology Solutions Group\\ Intel Corporation, Haifa, Israel\\
\url{alexander.nadel@intel.com}}

\maketitle
\thispagestyle{empty}

\begin{abstract}
The task of extracting an unsatisfiable core for a given Boolean formula has 
been finding 
more and more applications in recent years. 
The only existing approach that scales well for large real-world formulas 
exploits the ability 
of modern SAT solvers to produce resolution refutations. 
However, the resulting  unsatisfiable cores are suboptimal. 
We propose a new algorithm for minimal unsatisfiable core extraction, 
based on a deeper exploration of resolution-refutation properties. 
We provide experimental results on formal verification benchmarks confirming that our algorithm finds smaller cores than suboptimal algorithms; and that it runs faster than those algorithms
that guarantee minimality of the core.
\end{abstract}

\section{Introduction}
Many real-world problems, arising in formal verification of hardware and 
software, planning and other areas, 
can be formulated as constraint satisfaction problems, 
which can be translated into Boolean formulas in conjunctive normal form (CNF). 
Modern Boolean satisfiability (SAT) solvers, such as 
Chaff~\cite{Chaff2001,Chaff2004} 
and MiniSAT~\cite{DBLP:conf/sat/EenS03}, which implement enhanced versions of 
the 
Davis-Putnam-Longeman-Loveland (DPLL) backtrack-search algorithm, 
are often able to determine whether a large formula is satisfiable or 
unsatisfiable. 
When a formula is unsatisfiable, it is often required to find an 
\emph{unsatisfiable core}---that is, 
a small unsatisfiable subset of the formula's clauses. 
Example applications include functional verification of hardware~\cite{FuncVer}, 
FPGA routing~\cite{FPGARouting}, and abstraction 
refinement~\cite{DBLP:conf/tacas/McMillanA03}. 
For example, in FPGA routing, an unsatisfiable instance implies that the channel 
is unroutable. 
Localizing a small unsatisfiable core is necessary to determine the underlying 
reasons for the failure. 
An unsatisfiable core is a \emph{minimal unsatisfiable core (MUC)}, 
if it becomes satisfiable whenever any one of its clauses is removed. 
It is always desirable to find a minimal unsatisfiable core, 
but this problem is very hard 
(it is $D^P$-complete; see~\cite{MUC-DPComplete}).

In this paper, we propose an algorithm that is able to find a minimal 
unsatisfiable core for large ``real-world'' formulas. Benchmark families, arising in 
formal verification of hardware (such as~\cite{velevbench}), are of particular 
interest for us. The only approach for unsatisfiable core extraction that scales 
well for formal verification benchmarks was independently proposed 
in~\cite{ECC_Zhang} and in~\cite{DBLP:conf/date/GoldbergN03}. We refer to this 
method as the \textit{EC (Empty-clause Cone)} algorithm. EC exploits the ability 
of modern SAT solvers to produce a resolution refutation, given an unsatisfiable 
formula. Most state-of-the-art SAT solvers, beginning with 
GRASP~\cite{GRASP} and including
Chaff~\cite{Chaff2001,Chaff2004} and MiniSAT~\cite{DBLP:conf/sat/EenS03},
implement a DPLL backtrack search enhanced by a failure-driven 
assertion loop~\cite{GRASP}. These solvers explore the assignment tree 
and create new \emph{conflict clauses} at the leaves of the tree, using 
resolution on the initial clauses and previously created conflict clauses. This 
process stops when either a satisfying assignment is found or when the empty 
clause ($\square$) is derived. In the latter case, SAT solvers are able to 
produce a \emph{resolution refutation}---a directed acyclic graph (dag), whose 
vertices are associated with clauses, and whose edges describe resolution 
relations between clauses. The sources of the refutation are the 
initial clauses and the empty clause $\square$ is a sink. EC traverses a 
reversed refutation, starting with $\square$ and taking initial clauses, 
connected to $\square$, as the unsatisfiable core. Invoking EC until a fixed 
point is reached~\cite{ECC_Zhang} allows one to reduce the unsatisfiable core 
even more. We refer to this algorithm as \textit{EC-fp}. However, the resulting 
cores can be reduced further.

The basic flow of the algorithm for 
minimal unsatisfiable core extraction proposed in this paper is
composed of the following steps: 
\begin{enumerate}
\item 
Produce a resolution refutation $\Pi$ of a given formula using a SAT 
solver. 
\item 
Drop from $\Pi$ all clauses not connected to $\square$. 
At this point, all the initial clauses remaining in $\Pi$ comprise an 
unsatisfiable core. 
\item 
For every initial clause $C$
remaining in $\Pi$, check whether it belongs to a minimal unsatisfiable core (MUC) in 
the following manner: 
\begin{quote}
Remove $C$ from $\Pi$, along with
all conflict clauses for which $C$ was required to derive them. Pass all the remaining clauses 
(including conflict clauses) to a SAT solver. 
\begin{itemize}
\item
If they are satisfiable, then $C$ belongs to a 
minimal unsatisfiable core, so continue with another initial clause.
\item
If the clauses are unsatisfiable, then
$C$ does not belong to a MUC, so  
replace $\Pi$ by a new valid resolution refutation not containing $C$. 
\end{itemize}
\end{quote}
\item
Terminate when all the initial clauses remaining in $\Pi$ 
comprise a MUC.
\end{enumerate}

Related work is discussed in the next section. Section~\ref{sec:rr} is 
dedicated to refutation-related definitions. Our basic \textit{Complete 
Resolution Refutation (CRR)} algorithm is described in Sect.~\ref{sec:crr},
and a 
pruning technique, enhancing CRR and called \emph{Resolution Refutation-based 
Pruning (RRP)}, is described in Sect.~\ref{sec:ecrr}. Experimental results are 
analyzed in Sect.~\ref{sec:er}. This is followed up by a brief conclusion.

\section{Related Work}
\label{sec:rw}
Algorithms for unsatisfiable core based on the ability of modern SAT solvers to 
produce resolution refutations~\cite{ECC_Zhang,DBLP:conf/date/GoldbergN03} are 
the most relevant for our purposes for two reasons. First, this approach allows 
one to deal with real-world examples arising in formal verification. Second, it 
serves as the basis of our algorithm. We have already described the EC and EC-fp 
algorithms in the introduction. Here we briefly consider other approaches.

Theoretical work (e.g., \cite{Th5}) has concentrated on developing efficient 
algorithms for formulas with small \emph{deficiency} (the number of clauses 
minus the number of variables). However, real-world formulas have arbitrary (and 
usually large) deficiency. A number of works considered the harder problem of 
finding the smallest minimal unsatisfiable core~\cite{LynceMinimum,BBMinimum}, or 
even finding all minimally unsatisfiable formulas ~\cite{AllMinUnsat}. As one
can imagine, these algorithms are not scalable for even moderately large real-world 
formulas.

In \cite{bruni01,bruni03}, an ``adaptive core search'' is applied for finding a 
small unsatisfiable core. The algorithm starts with a very small satisfiable 
subformula, consisting of \textit{hard} clauses. The unsatisfiable core is built 
by an iterative process that expands or contracts the current core by a fixed 
percentage of clauses. The procedure succeeds in finding small, though not 
necessarily minimal, unsatisfiable cores for the problem instances it was tested on,
but these are very small and artificially generated.

Another approach that allows for finding small, but not necessarily minimal, 
unsatisfiable cores is called AMUSE~\cite{AMUSE}. In this approach, selector variables 
are added to each clause and the unsatisfiable core is found by a branch-and-bound 
algorithm on the updated formula. Selector variables allow the program to 
implicitly search for unsatisfiable cores using an enhanced version of DPLL on 
the updated formula. The authors note their methods ability to locate different 
unsatisfiable cores, as well as its inability to cope with large formulas.

The above described algorithms do not guarantee minimality of the 
cores extracted. One folk algorithm for minimal unsatisfiable core 
extraction, which we dub \textit{Na\"{\i}ve}, works as follows: For 
every clause $C$ in an unsatisfiable 
formula $F$, Na\"{\i}ve checks if it belongs to the 
minimal unsatisfiable core, by invoking a SAT solver on $F 
\setminus C$.  Clause $C$ does not belong to MUC if and only if the 
solver finds that $F \setminus C$ is unsatisfiable, in which case $C$ is 
removed from $F$. In the end, $F$ contains a minimal unsatisfiable core.

The only non-trivial algorithm existing in the current literature that guarantees 
minimality is MUP~\cite{MUP}. MUP is mainly a prover of minimal 
unsatisfiability, as opposed to an unsatisfiable core extractor. It decides the 
minimal unsatisfiability of a CNF formula through BDD manipulation. When MUP is 
used as a core extractor, it removes one clause at a time until the remaining 
core is minimal.  MUP is able to prove minimal unsatisfiability of some 
particularly hard classical problems quickly, whereas even just proving 
unsatisfiability is a challenge for modern SAT solvers. However, 
the formulas described in~\cite{MUP} are small and arise in 
areas other than formal verification. We will see in 
Section~\ref{sec:er} that MUP is significantly outperformed by Na\"{\i}ve on formal 
verification benchmarks.

\section{Resolution Refutations}
\label{sec:rr}
We begin with a resolution refutation of a given unsatisfiable formula,
defined as follows:

\begin{dfn}[Resolution refutation]
\label{fig:def_RRG}
Let $F$ be an unsatisfiable CNF formula (set of clauses)
and let $\Pi(V,E)$ be a dag whose 
vertices are clauses.\footnote{From now on, we shall use the terms ``vertex'' and 
``clause'' interchangeably in the context of resolution refutation.} Suppose $V = 
V^i \cup V^c$, where $V^i$ are all the sources of $\Pi$, referred to as 
\emph{initial clauses}, and $V^c=C^c_1,\ldots,C^c_m$ is an ordered set of 
non-source vertices, referred to as \emph{conflict clauses}. Then,
the dag $\Pi(V,E)$ 
is a \emph{resolution refutation of $F$} if:
\begin{enumerate}
    \item $V^i=F$;
    \item for every conflict clause $C^c_i$, there exists a resolution 
derivation $\{D_1, D_2,\ldots, D_k, C^c_i\}$, such that:

\begin{enumerate}
    \item for every $j =1,\ldots,k$, $D_j$ is either an initial clause or a 
prior conflict clause $C^c_f$, $f < i$, and
    \item \label{fig:def_edges} there are edges $D_1\rightarrow C^c_i,\ldots,D_k\rightarrow C^c_i\in E$ 
    (these are the only edges in $E$);
\end{enumerate}

    \item the sink vertex $C^c_m$ is the only empty clause in $V$.
\end{enumerate}
\end{dfn}

For the subsequent discussion, it will be helpful to capture the notion of 
vertices that are ``reachable'', or ``backward reachable'', from a given clause in a given dag.

\begin{dfn}[Reachable vertices]
\label{fig:def_rve}
Let $\Pi$ be a dag. A vertex $D$ is \emph{reachable} from $C$ if there is a 
path (of 0 or more edges) from $C$ to $D$. The set of all vertices reachable 
from $C$ in $\Pi$ is denoted $Re(\Pi, C)$. The set of all vertices 
unreachable from $C$ in $\Pi$ is denoted by $\overline{Re}(\Pi, C)$
\end{dfn}

\begin{dfn}[Backward reachable vertices]
\label{fig:def_brve}
Let $\Pi$ be a dag. A vertex $D$ is \emph{backward reachable} from $C$ if 
there is a path (of 0 or more edges) from $D$ to $C$. The set of all vertices 
backward reachable from $C$ in $\Pi$ is denoted by $\mbox{\it BRe} (\Pi, C)$. 
The set of all vertices not backward reachable from $C$ in $\Pi$ is denoted 
$\overline{\mbox{\it BRe}}(\Pi, C)$.
\end{dfn}

For example, consider the resolution refutation of Fig.~\ref{fig:muc_fig1}. We 
have $Re(\Pi, C^i_5)=\{C^i_5,C^c_2,C^c_3,C^c_4,C^c_5\}$ and $\mbox{\it BRe}(\Pi, 
C^c_4)=\{C^c_4,C^i_5,C^i_6\}$.

Resolution refutations trace all resolution derivations of conflict clauses, 
including the empty clause. Generally, not all clauses of a refutation are 
required to derive $\square$, but only such that are backward reachable 
from $\square$. It is not hard to see that even if all other clauses and related 
edges are omitted, the remaining graph is still a refutation. We refer to such 
refutations as \emph{non-redundant} (see Definition~\ref{fig:def_nrs_RRG}). The 
refutation in Fig.~\ref{fig:muc_fig1} is non-redundant.

To retrieve a non-redundant subgraph of a refutation, it is sufficient to take 
$\mbox{\it BRe} (\Pi, \square)$ as the vertex set and to restrict the edge set 
$E$ to edges having both ends in $\mbox{\it BRe} (\Pi, \square)$. We 
denote a non-redundant subgraph of a refutation $\Pi$ by 
$\Pi\!\upharpoonright_{\mbox{\it BRe}(\Pi,\square)}$. Observe that 
$\Pi\!\upharpoonright_{\mbox{\it BRe}(\Pi,\square)}$ is a valid non-redundant refutation.

\begin{dfn}[Non-redundant resolution refutation]
\label{fig:def_nrs_RRG}
A resolution refutation $\Pi$ is 
\emph{non-redundant} if there is a path in $\Pi$ from every clause to $\square$.
\end{dfn}

Lastly, we define a relative hardness of a resolution refutation.

\begin{dfn}[Relative hardness]
\label{fig:def_relsize}
The \emph{relative hardness} of 
a resolution refutation is the ratio between the total number of clauses and the number 
of initial clauses.
\end{dfn}

\section{The Complete Resolution Refutation (CRR) Algorithm}
\label{sec:crr}
Our goal is to find a minimal unsatisfiable core of a given unsatisfiable 
formula $F$. The proposed \emph{CRR} method is displayed as Algorithm~\ref{fig:muc1}.

\begin{algorithm}
\caption{{\bf (CRR).} Returns a MUC, given an unsatisfiable formula $F$.}
\label{fig:muc1}
\begin{algorithmic}[1]
\STATE Build a non-redundant refutation $\Pi(V^i \cup V^c,E)$
\WHILE{unmarked clauses exist in $V^i$}
\STATE \label{lbl:pick} $C \leftarrow $ \textit{PickUnmarkedClause($V^i$)}
\STATE Invoke a SAT solver on $\overline{Re}(\Pi, C)$
\IF{$\overline{Re}(\Pi, C)$ is \textit{satisfiable}}
\STATE \label{lbl:sat} Mark $C$ as a MUC member
\ELSE
\STATE \label{lbl:unsat1} Let $G = \overline{Re}(\Pi, C)$
\STATE Build resolution refutation $\Pi'(V^i_G \cup V^c_G, E_G)$
\STATE $V^i \leftarrow V^i \setminus \{ C \}$
\STATE $V^c \leftarrow (V^c \setminus Re(\Pi, C)) \cup V^c_G$
\STATE $E \leftarrow (E \setminus Re^E(\Pi, C)) \cup E_G$
\STATE \label{lbl:unsat2} $\Pi(V^i \cup V^c,E) \leftarrow \Pi(V^i \cup 
V^c,E)\upharpoonright_{\mbox{\it BRe}(\Pi,\square)}$
\ENDIF
\ENDWHILE
\RETURN $V^i$
\end{algorithmic}
\end{algorithm}

First, CRR builds a non-redundant resolution refutation. Invoking a SAT solver 
for constructing a (possibly redundant) resolution refutation $\Pi(V,E)$ and 
restricting it to $\Pi\!\upharpoonright_{\mbox{\it BRe}(\Pi,\square)}$ is 
sufficient for this purpose.

Suppose $\Pi(V^i \cup V^c,E)$ is a non-redundant refutation. CRR checks, for 
every unmarked clause $C$ left in $V^i$, whether $C$ belongs to the minimal 
unsatisfiable core. Initially, all clauses are unmarked. At each stage of the 
algorithm, CRR maintains a valid refutation of $F$.

Recall from Definition~\ref{fig:def_rve} that $\overline{Re}(\Pi, C)$ is the 
set of all vertices  in $\Pi$ unreachable from $C$. By construction of 
$\Pi$, the $\overline{Re}(\Pi, C)$ clauses were derived independently of $C$. To 
check whether $C$ belongs to the minimal unsatisfiable core, we provide the SAT 
solver with $\overline{Re}(\Pi, C)$, including the conflict clauses. We are 
trying to \emph{complete the resolution refutation}, not using $C$ as one of the 
sources. Observe that $\square$ is always reachable from $C$, since $\Pi$ is a 
non-redundant refutation; thus $\square$ is never passed as an input to the SAT 
solver. We let the SAT solver try to derive $\square$, using 
$\overline{Re}(\Pi, C)$ as the input formula, or else prove that 
$\overline{Re}(\Pi, C)$ is satisfiable.

In the latter case, we conclude that $C$ must belong to the minimal 
unsatisfiable core, since we found a model for an unsatisfiable subset of 
initial clauses minus $C$. Hence, if the SAT solver returns 
\textit{satisfiable}, the algorithm marks $C$ (line~\ref{lbl:sat}) and moves to 
the next initial clause. However, if the SAT solver returns 
\textit{unsatisfiable}, we cannot simply remove $C$ from $F$ and move to the 
next clause, since we need to keep a valid resolution refutation 
for our algorithm to work properly. We describe the construction of a valid 
refutation (lines~\ref{lbl:unsat1}--\ref{lbl:unsat2}) next.

Let $G = \overline{Re}(\Pi, C)$. The SAT solver produces a new resolution 
refutation $\Pi'(V^i_G \cup V^c_G, E_G)$ for $G$, whose sources are the clauses 
$\overline{Re}(\Pi, C)$. We cannot use $\Pi'$ as the refutation for the 
subsequent iterations, since the sources of the refutation may only be initial 
clauses of $F$. However, the ``superfluous'' sources of $\Pi'$ are conflict 
clauses of $\Pi$, unreachable from $C$, and thus are derivable from $V^i 
\setminus C$ using resolution relations, corresponding to edges of $\Pi$. 
Hence, it is sufficient to augment $\Pi'$ with such edges of $\Pi$ that 
connect $V^i \setminus C$ and $\overline{Re}(\Pi, C)$ to obtain a 
valid refutation whose initial clauses belong to $F$. Algorithm CRR 
constructs a new refutation, whose sources are $V^i \setminus C$; the conflict 
clauses are $\overline{Re}(\Pi, C) \cup V^c_G$ and the edges are $(E \setminus 
(V_1,V_2)|(V_1 \in Re(\Pi, C)\ or\ V_2 \in Re(\Pi, C))) \cup E_G$. This new 
refutation might be redundant, since $\Pi'(V^i_G \cup V^c_G, E_G)$ is not 
guaranteed to be non-redundant. Therefore, prior to checking the next clause, we 
reduce the new refutation to a non-redundant one. Observe that in the process of 
reduction to a non-redundant subgraph, some of the initial clauses of $F$, may 
be omitted; hence, each time a clause $C$ is found not to belong to the minimal 
unsatisfiable core, we potentially drop not only $C$, but also other clauses.

\begin{figure}[tb]
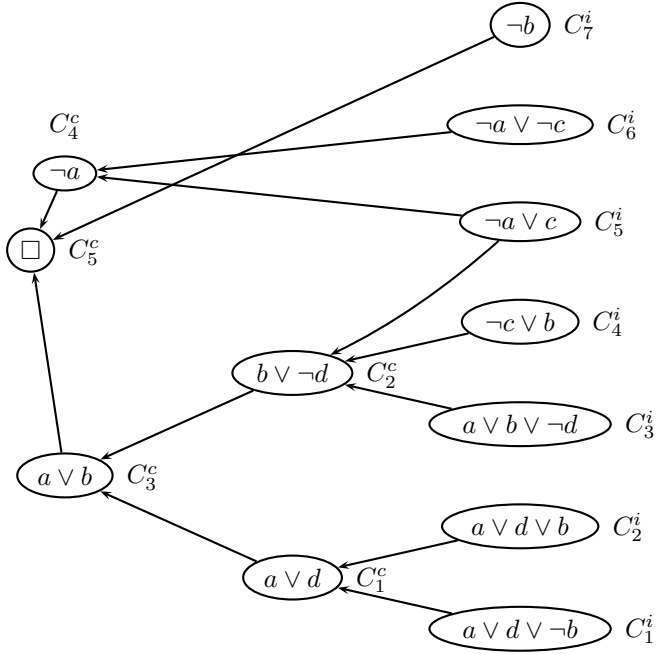

\psset{arrows=<-}
\pstree[treemode=R,angleB=180,levelsep=20ex,thislevelsep=3ex] 
{\Toval{$\square$}~[tnpos=r]{$C^c_5$}} {
    \skiplevels{2}
    \Toval{$\lnot b$}~[tnpos=r]{$C^i_7$}
    \endskiplevels
    \pstree{\Toval{$\lnot a$}~[tnpos=a]{$C^c_4$}} {
        \skiplevel{\Toval{$\lnot a \lor \lnot c$}~[tnpos=r]{$C^i_6$}}
        \skiplevel{\Toval[name=i5]{$\lnot a \lor c$}~[tnpos=r]{$C^i_5$}}
    }
    \pstree{\Toval{$a \lor b$}~[tnpos=r]{$C^c_3$}} {
        \pstree{\Toval[name=c2]{$b \lor \lnot d$}~[tnpos=r]{$C^c_2$}} {
            \Toval{$\lnot c \lor b$}~[tnpos=r]{$C^i_4$}
            \Toval{$a \lor b \lor \lnot d$}~[tnpos=r]{$C^i_3$}
        }
        \pstree{\Toval{$a \lor d$}~[tnpos=r]{$C^c_1$}} {
            \Toval{$a \lor d \lor b$}~[tnpos=r]{$C^i_2$}
            \Toval{$a \lor d \lor \lnot b$}~[tnpos=r]{$C^i_1$}
        }
    }
}
\ncarc{->}{i5}{c2}
\caption{Resolution refutation example}
\label{fig:muc_fig1}
\end{figure}

We demonstrate the process of completing a refutation on the example from 
Fig.~\ref{fig:muc_fig1}. Suppose we are checking whether $C^i_1$ belongs to 
the minimal unsatisfiable core. In this case, $G = \overline{Re}(\Pi, C^i_1) = 
\{C^i_2,C^i_3,C^i_4,C^i_5,C^i_6,C^i_7,C^c_2,C^c_4\}$. The SAT solver receives 
$G$ as the input formula. It is not hard to check that $G$ is unsatisfiable. One 
refutation of $G$ is $\Pi'(V^i_G \cup V^c_G, E_G)$, where $V^i_G=\{C^i_2, 
C^c_2, C^i_7, C^c_4\}$, $V^c_G=(D_1 = \square, D_2=a \lor b$), and $E_G$ 
$=\{(C^i_2,D_2),(C^c_2,D_2),(D_2,D_1),(C^i_7,D_1),(C^c_4,D_1)\}$.  Therefore, 
$C^i_1$, $C^c_1$, $C^c_3$, $C^c_5$ and related edges are excluded from the 
refutation of $F$, whereas $D_2$, $D_1$ and related edges are added to the 
refutation of $F$. In this case, the resulting refutation is non-redundant.

We did not define how the function \textit{PickUnmarkedClause} should pick 
clauses (line~\ref{lbl:pick}). Our current implementation picks clauses in the 
order in which clauses appear in the given formula. Development of 
sophisticated heuristics is left for future research. 

Another direction that may 
lead to a speed-up of CRR is adjusting the SAT solver for the purposes 
of CRR algorithm, considering that the SAT solver is invoked thousands of times on rather 
easy instances. Integrating the data structures of CRR and the SAT solver, 
tuning SAT solver's heuristics for CRR, and holding the refutation in-memory, rather 
than on disk (as suggested in~\cite{ECC_Zhang} for EC), 
can be helpful.

\begin{figure}[tb]
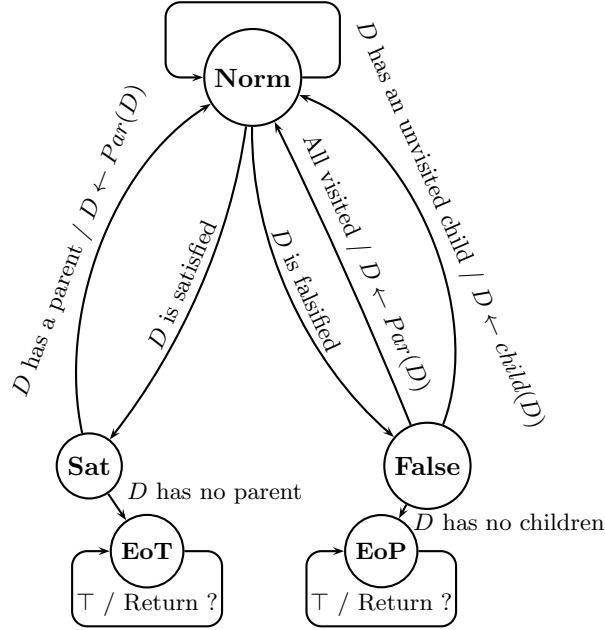

\centering
\vspace*{1cm}
\psmatrix[mnode=circle,colsep=30pt,rowsep=110pt]
& [name=Norm] {\bf Norm} \\ {\bf Sat} &  & [name=False]{\bf False} \\
\endpsmatrix
\psset{shortput=nab,arrows=->,labelsep=3pt}
\small
\ncarc[arcangle=15]{Norm}{2,1}\bput{:D}{$D$ is satisfied}
\ncloop[angleB=180,loopsize=1,arm=.5,linearc=.2]{Norm}{Norm}\bput{:D}{$D$ is not 
satisfied nor falsified / Return an unassigned literal}
\ncarc[arcangle=35]{2,1}{Norm}\aput{:U}{$D$ has a parent / $D \leftarrow 
Par(D)$}

\psmatrix[mnode=circle,colsep=30pt,rowsep=20pt]
[name=EOT]{\bf EoT} &  & [name=EOP]{\bf EoP} \\
\endpsmatrix
\ncline{2,1}{EOT}^{$D$ has no parent}
\ncarc[arcangle=-25]{Norm}{False}\aput{:U}{$D$ is falsified}
\ncarc[arcangle=-45]{False}{1,2}\bput{:D}{$D$ has an unvisited child / $D 
\leftarrow child(D)$}
\ncarc[arcangle=-2]{False}{1,2}\bput{:D}{All visited / $D \leftarrow 
Par(D)$}
\ncline{False}{EOP}^{$D$ has no children}
\ncloop[angleB=180,loopsize=-1,arm=.5,linearc=.2]{EOT}{EOT}\bput{:D}{$\top$ / 
Return ?}
\ncloop[angleB=180,loopsize=-1,arm=.5,linearc=.2]{EOP}{EOP}\bput{:D}{$\top$ / 
Return ?}
\bigskip
\caption{Function RRP\_Decide, represented as a transition relation. This 
function is invoked by the decision engine of a SAT solver, implementing the RRP 
pruning technique.}
\label{fig:eCRR_dec}
\end{figure}


\section{Resolution-Refutation-Based Pruning}\label{sec:ecrr}

In this section, we propose an enhancement of Algorithm CRR by 
developing resolution refutation-based pruning techniques for when a SAT solver
invoked on $\overline{Re}(\Pi, C)$ to check whether it is possible to complete 
a refutation without $C$. We refer to the pruning technique, proposed 
in this section, \emph{Resolution Refutation-based Pruning (RRP)}. We presume
that the reader is familiar with the functionality of a modern SAT solver. (An 
overview is given in~\cite{NadelThesis}.)

An assignment $\sigma$ \emph{falsifies} a clause $C$, if every literal of $C$ is 
\emph{false} under $\sigma$. An assignment $\sigma$ \emph{falsifies} a set of clauses 
$P$ if every clause $C \in P$ is falsified by $\sigma$. We claim that a model 
for $\overline{Re}(\Pi, C)$ can only be found under such a partial assignment 
that falsifies every clause in some path from $C$ to the empty clause in 
$Re(\Pi, C)$. The intuitive reason is that every other partial assignment 
satisfies $C$ and must falsify $\overline{Re}(\Pi, C)$, since $F$ is 
unsatisfiable. A formal statement and proof is provided in Theorem~\ref{theo:uu} below.

Consider the example of Fig.~\ref{fig:muc_fig1}. Suppose the currently 
visited clause is $C^i_5$. There exist two paths from $C^i_5$ to the empty 
clause $C^c_5$, namely $\{C^i_5,C^c_4,C^c_5\}$ and 
$\{C^i_5,C^c_2,C^c_3,C^c_5\}$. A model for $\overline{Re}(\Pi, C^i_5)$ can 
only be found in a subspace under the partial assignment $\{a=1,c=0\}$, falsifying 
all the clauses of the first path. The clauses of the second path cannot be 
falsified, since $a$ must be \emph{true} to falsify clause $C^i_5$ and \emph{false} to 
falsify clause $C^c_3$.

Denote a subtree connecting $C$ and $\square$ by 
$\Pi\!\upharpoonright_C$. The proposed pruning technique, RRP, is 
integrated into the decision engine of the SAT solver. The solver receives 
$\Pi\!\upharpoonright_C$, together with the input formula 
$\overline{Re}(\Pi, C)$. The decision engine of the SAT solver explores 
$\Pi\!\upharpoonright_C$ in a depth-first manner, picking unassigned variables 
in the currently explored path as decision variables and assigning them 
\textit{false}. As usual, Boolean Constraint Propagation (BCP) follows each 
assignment. Backtracking in $\Pi\!\upharpoonright_C$ is tightly related 
to backtracking in the assignment space. Both happen when a satisfied clause in 
$\Pi\!\upharpoonright_C$ is found or when a new conflict clause is 
discovered during BCP. After a particular path in $\Pi\!\upharpoonright_C$ 
has been falsified, a general-purpose decision heuristic is used until the SAT 
solver either finds a satisfying assignment or proves that no such assignment 
can be found under the currently explored path. This process continues until 
either a model is found or the decision engine has completed exploring 
$\Pi\!\upharpoonright_C$. In the latter case, one can be sure that no 
model for $\overline{Re}(\Pi, C)$ exists. However, the SAT solver should 
continue its work to produce a refutation. 

We need to describe in greater detail the changes in 
the decision and conflict analysis engines of the SAT solver required to implement 
RRP.
The decision engine first invokes function \textit{RRP\_Decide}, depicted in 
Fig.~\ref{fig:eCRR_dec}, as a state transition relation. Each transition 
edge has a label consisting of a condition under which the state transition 
occurs and an operation, executed upon transition. The state can be one of the following:
\begin{center}\begin{tabular}{l@{\qquad}l}
({\bf Norm}) & normal;\\
({\bf Sat}) & the currently explored clause is satisfied;\\
({\bf False}) & the currently explored clause is falsified;\\
({\bf EoT}) & subgraph $\Pi\!\upharpoonright_C$ has been explored;\\
({\bf EoF}) & all clauses in the currently explored path are falsified.\\
\end{tabular}\end{center}
The 
states are managed globally, that is, if \textit{RRP\_Decide} moves to state $S$, it 
will start in state $S$ when invoked next. A pointer $D$ to the 
currently visited clause of $\Pi\!\upharpoonright_C$ is also managed 
globally. The state transition relation is initialized prior to the first 
invocation of the decision engine. Pointer $D$ is initialized to $C$ and the 
initial state is {\bf Norm}.

State {\bf Norm} corresponds to a situation when the algorithm does not know what
the status of $D$ is. If $D$ is neither satisfied nor falsified, \textit{RRP\_Decide} returns 
a negation of some literal of $D$, which will serve as the next decision 
variable. If $D$ is satisfied, the algorithm moves to {\bf Sat}. Observe that a 
clause may become satisfied only as a result of BCP. 
Encountering a satisfied clause means that the currently explored path cannot be 
falsified, and we can backtrack. Suppose we are in {\bf Sat}, meaning that $D$ is 
satisfied. If $D$ has a parent, the algorithm backtracks by moving $D$ to point 
to its parent, and goes back to {\bf Norm}; otherwise, the tree is explored and 
the algorithm moves to {\bf EoT}. In this case, \textit{RRP\_Decide} returns an unknown 
value and a general-purpose heuristic must be used. Consider now the case 
when the state is {\bf Norm} and $D$ is falsified. The algorithm moves to {\bf False}. 
Here, one of the three conditions hold: 
\begin{itemize}
\item[(a)] $D$ has an unvisited child $S$. In 
this case $D$ is updated to point to $S$ and \textit{RRP\_Decide} moves back to  
{\bf Norm}. 
\item[(b)] All children of $D$ are visited. In this case, we backtrack by moving 
$D$ back to its parent and go back to {\bf Norm}.
\item[(c)] $D$ has no children. In this 
case, all the clauses in the currently explored path are falsified. The 
algorithm moves to {\bf EoF}; \textit{RRP\_Decide} returns an unknown value; and a 
general-purpose heuristic must be used.
\end{itemize}

To complete the picture, we describe the changes to the conflict analysis 
engine required to implement RRP. One of the main tasks of conflict analysis 
in modern SAT solvers is to decide to which level in the decision tree the algorithm 
should backtrack. Let this decision level be $bl$. When 
invoked in RRP mode, the conflict analysis engine must also find whether it is 
required to backtrack in $\Pi\!\upharpoonright_C$, and to which clause. 
The goal is to backtrack to the highest clause $B$ in the currently explored 
path in $\Pi\!\upharpoonright_C$, such that $B$ has unassigned literals. 
Recall that $D$ is a pointer to the currently visited clause of 
$\Pi\!\upharpoonright_C$. Denote by $mdl(D)$ the maximal decision 
level of $D$'s literals. If $bl \geq mdl(D)$, the algorithm does nothing; 
otherwise, it finds the first predecessor of $D$ in 
$\Pi\!\upharpoonright_C$, such that $bl < mdl(B)$ and sets $D \leftarrow 
B$.

We found experimentally that the optimal performance for RRP is achieved when 
it explores $\Pi\!\upharpoonright_C$ starting from $\square$ toward $C$ 
(and not vice-versa). In other words, prior to the search, the SAT solver 
reverses all the edges of $\Pi\!\upharpoonright_C$ and sets the pointer 
$D$ to $\square$, rather than to $C$. (By default, the current version of
RRP explores the graph only until 
a predefined depth of 50.) The next literal from the currently visited clause is 
chosen by preferring an unassigned literal with the maximal number of 
appearances in recent conflict clause derivations (similar to Berkmin's~\cite{Berkmin}
heuristic for SAT). The next visited child is chosen arbitrary. 
Further tuning of the algorithm is left for future research.

\begin{theo}
\label{theo:uu}
Let $\Pi(V^i,V^c)$ be a non-redundant resolution refutation. Let $C \in V^i$ 
be an initial clause and $\sigma$ be an assignment. Then, if $\sigma \models 
\overline{Re}(\Pi, C)$, there is a path $P = \{C,\ldots,C^c_m\}$ in $Re(\Pi, C)$, 
connecting $C$ to the empty clause\footnote{The empty clause always belongs to 
$Re(\Pi, C)$, since $\Pi(V^i,V^c)$ is non-redundant.}, such that 
$\sigma$ falsifies every clause in $P$.
\end{theo}

\begin{proof}
Suppose, on the contrary, that no such path exists. Then, there exists a satisfiable 
vertex cut $U$ in $\Pi$. But the empty clause is derivable from $U$, since it 
is a vertex cut; thus $U$ unsatisfiable, a contradiction.
\end{proof}

\begin{table}[tb]
\caption{Comparing algorithms for unsatisfiable core extraction. Columns \textbf{Instance}, \textbf{Var} and
\textbf{Cls} contain instance name, number of variables, and clauses, respectively. The next seven
columns contain execution times (in seconds) and core sizes (in number of clauses) for each algorithm. The cut-off time was 24 hours (86,400 sec.).
Column \textbf{Rel. Hard.} contains the relative hardness of the final
resolution refutation, produced by CRR+RRP.  Bold times are the best among algorithms guaranteeing minimality.}
\label{tab:results}
\centering\scriptsize
\begin{tabular}{ l || r | r || r | r || r | r | r | r | r || r}

\multirow{3}*{} & \multicolumn{1}{c|}{} & \multicolumn{1}{c||}{} & \multicolumn{2}{c||}{\textbf{Subopt.}} &
\multicolumn{2}{c|}{\textbf{CRR}} & \multicolumn{2}{c|}{\textbf{Na\"{\i}ve}} &
\multicolumn{1}{c||}{\textbf{MUP}} &  \multicolumn{1}{c}{\textbf{Rel.}} \\
                            \multicolumn{1}{c||}{\textbf{Instance}} &
\multicolumn{1}{c|}{\textbf{Var}} & \multicolumn{1}{c||}{\textbf{Cls}} &
\multicolumn{1}{c|}{\textbf{EC}} & \multicolumn{1}{c||}{\textbf{EC-fp}} &
\multicolumn{1}{c|}{\textbf{RRP}} & \multicolumn{1}{c|}{\textbf{plain}} &
\multicolumn{1}{c|}{\textbf{EC-fp}} & \multicolumn{1}{c|}{{\scriptsize\bf AMUSE}} &
\multicolumn{1}{c||}{\textbf{EC-fp}} &
\multicolumn{1}{c}{\textbf{Hard.}}\\
\hline\hline

\multirow{3}*{}\textsl{4pipe}   & 4237 &     & 9 & 171 & \textbf{3527} & 4933 &  24111 & time-out & time-out
&   1.4 \\
 & & 80213&23305 &17724 & 17184  & 17180 & 17182 &   &   &

 \\
\multirow{3}*{}\textsl{4pipe\_1\_ooo}   & 4647 &     & 10 & 332 & \textbf{4414} & 10944 & 25074 & time-out &
mem-out &            1.7\\
 & & 74554&24703 &14932 & 12553  & 12515 & 12374 &   &   &

 \\
\multirow{3}*{}\textsl{4pipe\_2\_ooo}   &   4941 &   & 13 & 347 & \textbf{5190} & 12284 & 49609 & time-out &
mem-out &    1.7\\
 & & 82207&25741 &17976 & 14259  & 14192 & 14017 &   &   &

 \\
\multirow{3}*{}\textsl{4pipe\_3\_ooo}   &   5233 &   & 14 & 336 & \textbf{6159} & 15867 & 41199 & time-out &
mem-out &    1.6\\
 & & 89473&30375 &20034 & 16494  & 16432 & 16419 &   &   &

 \\
\multirow{3}*{}\textsl{4pipe\_4\_ooo}   &   5525 &   & 16 & 341 & \textbf{6369} & 16317 & 47394 & time-out &
mem-out &    1.6\\
 & &96480 &31321 &21263 & 17712  & 17468 & 17830 &   &   &

 \\ \hline

\multirow{3}*{}\textsl{3pipe\_k}    & 2391 &     & 2 & 20 & \textbf{411} & 493 &    2147 & 12544    &
mem-out &    1.5\\
 & & 27405&10037 &6953 & 6788    & 6786  & 6784  & 6790  &   &

 \\
\multirow{3}*{}\textsl{4pipe\_k}    & 5095 &     & 8 & 121 & \textbf{3112} & 3651 &  15112 & time-out &
time-out &           1.5\\
 & &79489 &24501 &17149 & 17052  & 17078 & 17077 &   &   &

 \\
\multirow{3}*{}\textsl{5pipe\_k}    & 9330 &     & 16 & 169 &    \textbf{13836} & 17910 &    83402 & time-out
& mem-out &            1.4\\
 & & 189109&47066 &36571 & 36270  & 36296 & 36370 &   &   &

 \\ \hline
\multirow{3}*{}\textsl{barrel5} & 1407 &     & 2 & 19 &  93 & \textbf{86} & 406 &   326 & mem-out & 1.8\\
 & & 5383&3389  &3014 & 2653    & 2653  & 2653  & 2653  &   &

 \\
\multirow{3}*{}\textsl{barrel6} & 2306 &     & 35 & 322 &  \textbf{351} & 423 &    4099 &  4173 & mem-out &
1.8\\
 & & 8931&6151  &5033 & 4437    & 4437  & 4437  & 4437  &   &

 \\
\multirow{3}*{}\textsl{barrel7} & 3523 &     & 124 & 1154 & \textbf{970} & 1155 &   6213 &  24875 & mem-out &
1.9\\
 & & 13765&9252  &7135 & 6879    & 6877  & 6877  & 6877  &   &

 \\
\multirow{3}*{}\textsl{barrel8} & 5106 &     & 384 & 9660 & \textbf{2509} & 2859 &  time-out &  time-out &
mem-out &    1.8\\
 & & 20083&14416 &11249 & 10076  & 10075 & & &   &

 \\ \hline
%
\multirow{3}*{}\textsl{longmult4}   & 1966 &     & 0 & 0 &  8 & \textbf{7} &  109 & 152    & 13 &
2.6\\
 & & 6069&1247  &1246 & 972 & 972   & 972   & 976   &972    &

 \\
\multirow{3}*{}\textsl{longmult5}   & 2397 &     & 0 & 1 &  74 & \textbf{31} &   196 &   463 & 35 &
3.6\\
 & & 7431&1847  &1713 & 1518    & 1518  & 1518  & 1528  &1518   &

 \\
\multirow{3}*{}\textsl{longmult6}   & 2848 &     & 2 & 13 &  \textbf{288} & 311 &    749 &   2911 & 5084 &
5.6\\
 & & 8853&2639  &2579 & 2187    & 2187  & 2187  & 2191  &2187   &

 \\
\multirow{3}*{}\textsl{longmult7}   & 3319 &     & 17 & 91 & 6217 & \textbf{3076} & 6154 &  32791 & 68016 &
14.2\\
 & & 10335&3723  &3429 & 2979    & 2979  & 2979  & 2993  &2979   &
 \\
 \hline
\end{tabular}\end{table}


\section{Experimental Results}\label{sec:er}

We have implemented CRR and RRP in the framework of the VE solver. VE, a 
variant of the industrial solver Eureka, is a modern SAT solver, which 
implements the following state-of-the-art algorithms and techniques for SAT: it 
uses 1UIP conflict clause recording~\cite{Chaff2001}, enhanced by conflict 
clause minimization~\cite{MiniSAT}, frequent search 
restarts~\cite{Berkmin,Chaff2004}, an aggressive clause deletion 
strategy~\cite{Berkmin,Chaff2004}, and decision stack 
shrinking~\cite{NadelThesis,Chaff2004}. VE uses Berkmin's decision 
heuristic~\cite{Berkmin} until 4000 conflicts are detected, and then switches to the CBH 
heuristic, described in~\cite{CBH}. We used benchmarks from four well-known unsatisfiable 
families, taken from bounded model checking (\textsl{barrel}, 
\textsl{longmult})~\cite{mcbench} and microprocessor verification (\textsl{fvp-unsat.2.0, pipe\_unsat\_1.0})~\cite{velevbench}. All the instances we used appear in the first 
column of Table~\ref{tab:results}. 
The experiments on families \textsl{barrel} and \textsl{fvp-unsat.2.0} were carried out on a 
machine with 4Gb of memory and two Intel Xeon CPU 3.06 processors. A machine 
with the same amount of memory and two Intel Xeon CPU 3.20 processors was used 
for experiments with the families \textsl{longmult} and \textsl{pipe\_unsat\_1.0}.

Table~\ref{tab:results} summarizes the results of a comparison of the performance
of two algorithms for suboptimal unsatisfiable core extraction and five algorithms for minimal unsatisfiable core extraction in terms of execution time and core sizes.

First, we compare algorithms for minimal unsatisfiable core extraction, namely, Na\"{\i}ve, MUP, plain CRR, and CRR enhanced by RRP. In preliminary 
experiments, we found that Na\"{\i}ve demonstrates its best performance on 
formulas that are first trimmed down by a suboptimal algorithm for unsatisfiable core 
extraction. We tried Na\"{\i}ve in combination with EC, EC-fp and AMUSE and 
found that EC-fp is the best front-end for Na\"{\i}ve. In our main experiments, 
we used Na\"{\i}ve, combined with EC-fp, and Na\"{\i}ve combined with AMUSE. We 
have also found that MUP demonstrates its best performance when combined with 
EC-fp, while CRR performs the best when the first refutation is 
constructed by EC, rather than EC-fp. Consequently, we provide results for MUP 
combined with EC-fp and CRR combined with EC. MUP requires a so-called 
``decomposition tree'', in addition to the CNF formula. We used the c2d 
package~\cite{c2d} for decomposition tree construction.

The sizes of the cores do not vary much between MUC algorithms, so we concentrate on a performance comparison. One can see that the combination of EC-fp and Na\"{\i}ve outperforms the combination of AMUSE and Na\"{\i}ve, as well as MUP. Plain CRR outperforms Na\"{\i}ve on every benchmark, whereas CRR+RRP
outperforms Na\"{\i}ve on 15 out of 16 benchmarks (the exception being the hardest instance of \textsl{longmult}). This demonstrates that our
algorithms are justified practically. Usually, the
speed-up of these algorithms over Na\"{\i}ve varies between 4
and 10x, but it can be as large as 34x (for the hardest instance of \textsl{barrel} family)
and as small as 2x (for the hardest instance of \textsl{longmult}). RRP improves performance on most instances. The most significant speed-up of RRP is about 2.5x, achieved on hard instances of Family \textsl{fvp-unsat.2.0}. The only family for which RRP is usually unhelpful is \textsl{longmult}.

A natural question is why the complex instances of family \textsl{longmult} are hard for CRR, and even harder 
for RRP. The key difference between \textsl{longmult} and  other families is the 
hardness of the resolution proof. The relative hardness of a resolution 
refutation produced by CRR+RRP varies between 1.4 to 2 for every instance of 
every family, except \textsl{longmult}, where it reaches 14.2 for the \textsl{longmult7} instance. 
When the refutation is too complex, the exploration of $\overline{Re}(\Pi, 
C)$ executed by RRP is too complicated; thus, plain CRR is advantageous over 
CRR+RRP. Also, when the refutation is too complex, it is costly to perform 
traversal operations, as required by CRR. This explains why the advantage of CRR over 
Na\"{\i}ve is as small as 2X.

Comparing CRR+RRP on one side and EC and EC-fp on the other,
we find that CRR+RRP always produce smaller cores than both EC and EC-fp. The average gain on all instances of cores produced by CRR+RRP over cores produced by EC and EC-fp is 53\% and 11\%, respectively. The biggest average gain of CRR+RRP over EC-fp is achieved on Families \textsl{fvp-unsat.2.0} and \textsl{longmult} (18\% and 17\%, respectively). Unsurprisingly, both EC and EC-fp are usually much faster than CRR+RRP. However, on the three hardest instances of the barrel family, CRR+RRP outperforms EC-fp in terms of execution time.

\section{Conclusions}\label{sec:cfw}
We have proposed an algorithm for minimal unsatisfiable core extraction. It 
builds a resolution refutation using a SAT solver and finds a first 
approximation of a minimal unsatisfiable core. Then it checks, for every 
remaining initial clause $C$, if it belongs to the minimal unsatisfiable core. 
The algorithm reuses conflict clauses and resolution relations throughout its 
execution. We have demonstrated that our algorithm is faster than currently 
existing algorithms by a factor of 6 or more on large problems with non-overly hard
resolution proofs, and that it can find minimal unsatisfiable cores for real-world 
formal verification benchmarks.

\section*{Acknowledgments}
We thank Jinbo Huang for his help in obtaining and 
using MUP and Zaher Andraus for his help in receiving AMUSE.

\end{document}